\documentclass{amia}
\usepackage{graphicx}
\usepackage[numbers,super]{natbib}
\usepackage{algorithm}
\usepackage{algpseudocode}
\usepackage{amsfonts}
\usepackage{amsmath}
\usepackage{enumitem}
\usepackage{subcaption}
\usepackage{verbatim}
\usepackage{fancyvrb}
\usepackage{xcolor}
\usepackage{program}
\usepackage{float}
\usepackage{tablefootnote}
\usepackage{array,booktabs}

\begin{document}

\title{Secure Count Query on Encrypted Genomic Data}
\author{Mohammad Zahidul Hasan, Md Safiur Rahman Mahdi, Noman Mohammed}
\institutes{Department of Computer Science, University of Manitoba, Winnipeg, Manitoba, Canada}
\maketitle
\section*{Abstract}\setcounter{section}{0}
\textit{Capturing the vast amount of meaningful information encoded in the human genome is a fascinating research problem. The outcome of these researches have significant influences in a number of health related fields --- personalized medicine, paternity testing and disease susceptibility testing are a few to be named. To facilitate these types of large scale biomedical research projects, it oftentimes requires to share genomic and clinical data collected by disparate organizations among themselves. In that case, it is of utmost importance to ensure that sharing, managing and analyzing the data does not reveal the identity of the individuals who contribute their genomic samples. The task of storage and computation on the shared data can be delegated to third party cloud infrastructures, equipped with large storage and high performance computation resources. Outsourcing these sensitive genomic data to the third party cloud storage is associated with the challenges of the potential loss, theft or misuse of the data as the server administrator cannot be completely trusted as well as there is no guarantee that the security of the server will not be breached. In this paper, we provide a model for secure sharing and computation on genomic data in a semi-honest third party cloud server. The security of the shared data is guaranteed through encryption while making the overall computation fast and scalable enough for real-life large-scale biomedical applications. We evaluated the efficiency of our proposed model on a database of Single-Nucleotide Polymorphism (SNP) sequences and experimental results demonstrate that a query of 50 SNPs in a database of 50000 records, where each record contains 300 SNPs, takes approximately 6 seconds.}


\section{Introduction}
Analysis of human genome can reveal many essential information about an individual, like predisposition to a specific disease such as breast cancer, diabetes and Alzheimer~\cite{Mete2015Biomedical}. This analysis is usually done by querying an individual's genome against a list of known variations and then calculating the susceptibility \cite{Mete2015Biomedical}. To guarantee significant accuracy in this type of analysis, a large number of genomic sequences are required, the collection of which are sometimes beyond the capability of a sole organization~\cite{Burton2009Oxford}. Allowing the access of the genomic data surpassing the premise of the organization responsible for initial collection is a viable solution. But, handovering the access of the data, be it owned by a government organization or a private research institution, is not always very straightforward because of the nature of the genomic data.

Genomic data cannot be treated as any other data; it has some distinctive features. Naveed \textit{et al.} \cite{Naveed2015:ACM} identified six special features of genomic data. Genomic data does not change considerably over time and it is unique -- two individuals can easily be distinguished from their data. Furthermore, information about an individuals genotype, phenotype and blood relatives can also be inferred from his genomic data. Due to this sensitive nature, disclosure of this data has significant privacy risks. For example, a person carrying the mutation of a specific gene which increases the likelihood of developing a specific disease might be denied by an insurance company for his health coverage. Hence, while sharing these genomic data among multiple institutions, safety measures should be taken to uphold the privacy of the individuals who contributes the data. For this purpose, different privacy policies have been developed, thus facilitating the task of analysis to be done in a broader range. 

Genome-wide association study (GWAS) helps to understand and identify the generic variations that are associated with a particular disease. There are several type of variations that occur in human population, such as single-nucleotide polymorphism (SNP), copy-number variations (CNVs), rearrangement etc. GWAS produces aggregate statistics by examining these variations, typically SNPs from thousands of individuals which are used to determine the association between a SNP and a disease \cite{Yu2014GWAS}. Ensuring the security of the SNPs in this association studies is very important which has been clearly demonstrated by the work of Lin \textit{et al.} \cite{Lin2004Science} who showed that only 75 SNPs are enough to uniquely identify an individual. Besides, sensitive personal information can also be inferred from the aggregate statistics in GWAS \cite{Homer:2008:Plos, Wang:2009:ACM, Erlich2014nature}.

The volume of the aggregated shared data is enormous and requires vast amount of storage space. Due to the quality of services offered by the cloud infrastructures at considerably lower rate, especially having the characteristics of high availability and scalability, cloud computing services can be adopted for this purpose. But, cloud services are vulnerable to security threat and an adversary capable of breaching the security would be able to access the data residing into that server.  One published news clearly demonstrates that privacy should not be expected to be preserved from cloud service providers \cite{chen2009acm}.

In this paper, we aim at solving the security issues related to sharing and computation on outsourced genomic data. In particular, we address three potential security challenges. The first challenge is to guarantee \textit{data privacy}. The data stored in the cloud server, as well as the computation carried throughout the entire analysis process should be secured. Even if the cloud server gets compromised, the attacker should not learn anything about the data stored in the cloud. The second challenge is to provide \textit{query privacy}. The institutions contributing the data, the cloud service provider or an adversary should learn nothing about a query executed by a researcher or an institution. The third challenge is to achieve \textit{output privacy}. The result of the query should not be disclosed to anybody except the researcher who initiated the query.

Anonymization methods have been proved ineffective for protecting the genomic data \cite{Gymrek:2013:Science, Zhou:2011:ACM, MalinSweeney:2004} as these techniques incur high utility loss. Cryptographic techniques can compute a predefined function on encrypted dataset from multiple parties and returns the function's result without revealing any information about the data from different parties \cite{Erlich2014Plos}. For this reason, a number of privacy preserving techniques using cryptography have been developed to achieve the goal of sharing and computation on encrypted genomic data. However, none of these techniques can overcome the three challenges mentioned above simultaneously or scalable for real-life applications (see Section~\ref{Section:relatedWork} for more discussion). 

\textbf{Contributions.} In this paper, our goal is to design a framework for secure outsourcing of genomic data and securely computing count query on the outsourced data. Count query determines the number of matched records for a query predicate and it is very useful for genetic association studies to compute several statistical algorithms.

The main contributions of this paper are summarized below:
\begin{itemize}
\item We present a privacy preserving framework for secure count query operation. Genomic data is outsourced after encryption to a third party cloud server. Execution of a query is done by traversing an encrypted tree, where the decision of traversing each node is made by checking whether a query predicate matches with a particular branch of the tree. This checking is done through an interactive protocol using Yao's garbled circuit \cite{yao1986generate} between the query initiator (i.e., researcher) and the cloud server. Depending upon the query, branch(es) of the tree are traversed and the query result is calculated from the values stored in the nodes which match the query predicate. To determine the final output from the encrypted node values, we use \textit{Paillier cyptosystem} \cite{Paillier:1999:PCB}.
\item Our proposed model addresses all of the aforementioned three challenges, it provides -- \textit{data privacy, query privacy and output privacy} (the query result is only disclosed to the query initiator). Query is securely executed through an interactive protocol between the query initiator and the cloud server. The proposed method does not require an active participation of a trusted entity (i.e., proxy) for secure evaluation of the query or decryption of the result of the query.
\item We incorporate a tree-based indexing technique in our proposed model which not only provides an effective storage solution for large genomic datasets, but also queries can be executed efficiently by traversing the nodes of the tree. In addition, it facilitates an easy update operation. 

\item Through experiment and evaluation, we demonstrate the effectiveness of our approach. Experimental results clearly exhibit significant improvement in terms of time requirement for query evaluation in comparison with the previous approaches. For a count query operation over 40 SNPs in a database of 5000 records, it took approximately 30 min and 80 seconds for the methods proposed by Kantarcioglu \textit{et al.} \cite{Kantarcioglu2008IEEE} and Canim \textit{et al.} \cite{Canim2011IEEE}, respectively. The proposed approach significantly improves the runtime and it takes only 1.7 seconds to execute similar queries.
\end{itemize}

\section{Related work}\label{Section:relatedWork} 

\textbf{\begin{table}[t]
    \begin{center}
    \caption{Different properties of existing techniques for count query \label{table:related_work}}
    \begin{tabular}{ |c|c|c|c|c|c| }
        \hline
        \textbf{Algorithms} & \textbf{Method} & \textbf{Trusted Entity} & \multicolumn{3}{c|}{\textbf{Privacy}}\\\cline{4-6}
        \,&\,&\,& \textbf{Data} & \textbf{Query} & \textbf{Output}\\\hline
        Kantarcioglu \textit{et al.} \cite{Kantarcioglu2008IEEE} & HE & Online & \checkmark & \, & \,\\\hline
        Canim \textit{et al.} \cite{Canim2011IEEE} & Cryptographic Hardware, HE & N/A & \checkmark & \, & \checkmark\\\hline
        Our method & Pailler, GC & Offline & \checkmark & \checkmark & \checkmark\\\hline
    \end{tabular}
    \end{center}
\end{table}}
\textbf{Secure count query:} One of the earlier attempts that addressed the problem of secure outsourcing of genomic data for count query operation was a cryptographic model proposed by Kantarcioglu \textit{et al.} \cite{Kantarcioglu2008IEEE}. They proposed a framework which involves two different parties. One is responsible for integrating encrypted data coming from different data sources and then execute queries on behalf of a researcher on those encrypted data. Then the result of the query is sent to another party, a key holder site who is responsible for managing keys used for encryption and decryption. This key holder site decrypts the result and send the decrypted final result to the researcher. Their method does not provide query privacy; the cloud server receives the query in plaintext. Their method also does not provide output privacy as it is revealed to the key holder site. The main limitation of this work is efficiency. As mentioned earlier, it takes around 30 mins to execute a count query operation over 40 SNPs in a database of 5000 records. 

Canim \textit{et al.}~\cite{Canim2011IEEE} proposed a method using tamper-resistant cryptographic hardware to enable secure genomic data storage and processing at a single third party. The central server stores the encrypted genomic data coming from different sources and can compute over this data using the secure coprocessor (SCP) located at the server. Using SCP as cryptographic hardware and performing computation within the coprocessor is a secure approach for sharing biomedical data. However, it does not ensure query privacy; the query is issued in plaintext. On the other hand, it provides output privacy. The final output can be returned to the researcher by the SCP using a secure channel (e.g., SSL protocol). Hence, the cloud server can not see the output. The limitation of this approach is the cryptographic hardware itself. In partice, it is not always possible to ensure the availability of a secure coprocessor. In addition, coprocessor has a relatively small memory capacity and computational power \cite{Naveed2015:ACM}. So, it is not clear whether the method is scalable for big datasets \cite{Mete2015Biomedical}. 

To the best of our knowledge these are the only two works that have addressed the problem of secure count query operation. The comparison of our method with these two works is summarized in Table \ref{table:related_work}. The table summarizes different aspects of solving count query problem such as cryptographic methodology, involvement of trusted entity during query execution and different types of privacy offered by those methods. There are several other solutions that have been proposed to protect the privacy of both the outsourced data and the analysis. Although these works do not address the problem of secure count query operation, they target closely related problems and use different cryptographic techniques to ensure the security of the genomic data. Next we mention some of these works. 

\textbf{Other relevant works:} To protect the privacy of the genome database, Lauter \textit{et al.} \cite{Lauter2015:Springer} proposed a method where all the data are encrypted using \textit{fully homomorphic encryption} and then stored it in a single cloud server. They mainly focused on computing several statistical algorithms (e.g. \textit{Pearson Goodness-of-Fit} or \textit{Chi-Squared Test}, \textit{Linkage Disequilibrium}, \textit{Estimation Maximization (EM)} and \textit{Cochran-Armitage Test for Trend (CATT))} commonly used in genetic association studies. 
Kamm et al.~\cite{Kamm2013:Bioinformatics} employed secret sharing technique to guarantee the security of shared data and used secure multi-party computation to compute the value of different tests like \textit{${\chi}^2$ test}, \textit{Cochran-Armitage Test for Trend} and \textit{Transmission Disequilibrium Test}. Zhang \textit{et al.} \cite{Zhang2015Healer} used homomorphic encryption to compute \textit{Chi-Squared Test} in untrusted public cloud.

Cheon \textit{et al.} \cite{Cheon:2015:ED} adopted \textit{somewhat homomorphic encryption} scheme to ensure the security of the shared data in the cloud server. They only focused on computing the \textit{edit distance}. Zhang et al.~\cite{Zhang2015:BMC} used secret sharing and secure multi-party computation for computing \textit{edit distance} between two sequences. Wang et al. \cite{Wang:2015:ACM} also used the secure multi-party computation scheme to compute \textit{edit distance} to find similar patients based on the input from two different parties. Jha \textit{et al.} \cite{Jha:2008:ED} proposed a method to compute the similarity of DNA sequences using garbled circuits.

Ayday \textit{et al.} \cite{Ayday:2013:ACM} proposed a method for storing genomic data at a storage and processing unit and then processing it for medical tests and personalized medicine operations. The computations on this shared data are conducted using \textit{homomorphic encryption} and \textit{proxy re-encryption}. Yang \textit{et al.} \cite{Yang:2015:Hybrid} proposed a hybrid method for privacy preserving medical data sharing in cloud environment. They combined the existing privacy protecting ideas of \textit{privacy by statistics} and \textit{privacy by cryptography} and provided a hybrid search method which would be conducted across plaintext and ciphertext. Perl \textit{et al.} \cite{Perl:2012:IEEE} proposed a method for searching on biomedical data with a combination of \textit{Bloom filter} and \textit{homomorphic encryption}. They completely outsourced the task of searching in the third party cloud server. 

Xie \textit{et al.} \cite{Xie:2014:SecureMA} proposed a scheme for securely performing meta-analysis for genetic association study. Instead of storing data to multiple cloud storage (like Kamm \textit{et al.} ~\cite{Kamm2013:Bioinformatics} and Zhang \textit{et al.} ~\cite{Zhang2015:BMC}), they kept the data in the corresponding data owner's premises. Wang \textit{et al.} \cite{wang2016healer} designed a \textit{somewhat homomorphic encryption} based technique to compute \textit{exact logistic regression} to discover rare disease variants to analyze disease susceptibility in an untrusted cloud environment.

\section{Preliminaries}
In this section, we first present an overview of the system design and the format of the genomic dataset we consider in this paper. We then introduce the sample county query operation and threat model of our framework.

\subsection*{System Design Overview} 
\begin{figure}[t]
\centering
\captionsetup{justification=centering,margin=1cm}
    \includegraphics[width=.60\textwidth]{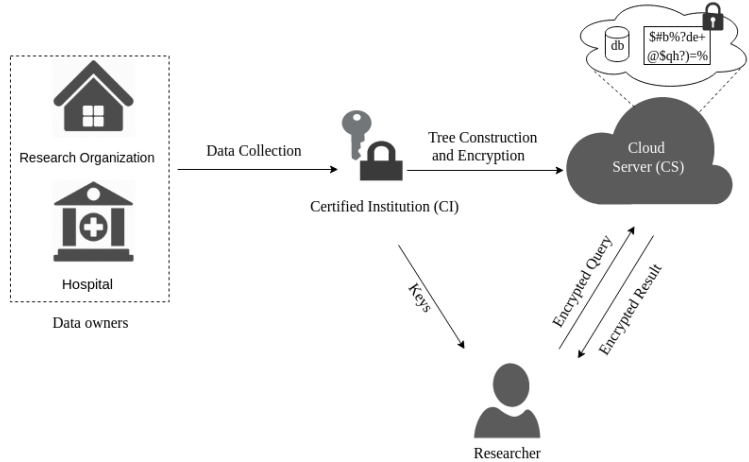}
    \caption{Proposed architecture of the system. 
    \label{fig:archi}}
\end{figure}

Figure \ref{fig:archi} presents a general overview of our proposed framework. As depicted in the figure, it incorporates four main participants: \textit{Data Owners}, \textit{Certified Institution (CI)}, \textit{Cloud Server (CS)} and \textit{Researchers}. Each entity is responsible for performing different specific tasks to make the overall system secure and functional. The roles performed by each of the entities are discussed below --
\begin{itemize}
\item \textit{\textbf{Data owners:}} \textit{Data owners} consists of the institutions who agreed to share the genomic data they possess. These institutions might be any academic institutions, non-academic research organizations, government research agencies or health departments such as the main contributors of data samples to dbGap  \cite{Paltoo:2014:Nature}. They send the genomic data to the \textit{CI} in plaintext. Prior to sending the data to the \textit{CI}, \textit{data owners} process their data in a formerly agreed format.
\item \textit{\textbf{Certified Institution (CI):}} The data shared by different \textit{data owners} reside in a database owned by a trusted entity which we call the \textit{CI}. Any trusted government institution such as National Institute of Health (NIH) in the United States can play this role. The responsibilities performed by \textit{CI} can be divided into two parts:\\
\textit{\textbf{1) Generation of index tree:}} 
Upon receiving the data from the contributing \textit{data owners}, \textit{CI} builds an encrypted searchable version of the aggregated shared data and sends it to the \textit{CS}. The search operation is basically performed on an encrypted \textit{index tree}. In our proposed system, the \textit{CI} builds only a single \textit{index tree} that contains all the records from the aggregated shared data and sends the encrypted version of the tree to the \textit{CS}. For any addition and deletion of records, \textit{CI} can update the tree accordingly in the \textit{CS}.\\
\textit{\textbf{2) Management of keys:}} Another responsibility of \textit{CI} is to manage the keys that the \textit{researchers} use to encrypt the query and decrypt the result of the query returned by the \textit{CS}. The data represented by the nodes of the \textit{index tree} are also encrypted using the same public key. 
\item \textit{\textbf{Cloud Server (CS):}} \textit{CS} gets the encrypted version of the \textit{index tree} and all the query operations are performed on this tree. \textit{CS} is responsible for handling all the communications with the \textit{researchers}. The \textit{researchers} send their encrypted query to \textit{CS}, \textit{CS} then executes this query and sends back the encrypted result to the \textit{researchers}.
\item \textit{\textbf{Researchers:}} \textit{Researchers} might be any individual or organization who is interested in performing query operations on the aggregated shared data residing in the \textit{CS}. To execute query on the outsourced data, \textit{researchers} need to obtain keys (both public and secret) from the  \textit{CI}. \textit{Researchers} use the public key to encrypt the query and sends it to the \textit{CS}. \textit{CS} evaluates this query on the encrypted tree and sends back the final result to the \textit{researchers}. Upon receiving the encrypted results, the \textit{researchers} use the secret key to decrypt the result. 
\end{itemize}

\subsection*{Genomic Data Representation}
Genome contains the hereditary information of an organism. The human genome is encoded in \textit{deoxyribonucleic acid} molecules which we commonly know as DNA. DNA molecules consist of two biopolymer chains each of which in turn consists of nucleotides. These nucleotides are represented as \textit{A, C, G, T} which are the acronyms of \textit{Adenine}, \textit{Cytosine}, \textit{Guanine} and \textit{Thymine} respectively. In DNA, these nucleotides form base pairs by making bonds with each other: \textit{A} bonds with \textit{T} and \textit{C} bonds with \textit{G}. There are approximately 3 billion base pairs in the whole human genomic sequence and most of them (99\%) are identical in two individuals. The remaining genomic variation distinguishes one individual from another. \textit{Single Nucleotide Polymorphism} (SNP) is a common DNA variation at a specific position in the genome which represents a difference in a single nucleotide. Most of the SNPs do not have any effect on human health. But some SNPs are directly responsible for developing a particular disease in the human body.

In this paper, we work with the databases consisting of SNP sequences. We assume that a sequence \textit{S} consists of multiple SNPs and we represent such a sequence as $S = \{a_1, a_2, ... ,a_n\}$ where $a_i$ represents a SNP. Table \ref{table:data_representation} represents an example of the format of the data that \textit{data owners} send to the \textit{CI}. Here, each row represents genomic sequences for one single patient. The SNPs $a_1, a_2, ... ,a_n$  are represented in each column. Each SNP can be represented by a pair of nucleotides and this is common in genomic data analysis \cite{Canim2011IEEE, purcell2007plink}. The last column indicates whether a genomic sequence is associated with cancer (positive) or not (negative). The dataset might contain other information about the phenotypes, but for keeping the structure of the data simple, we do not show those in Table \ref{table:data_representation}.

\textbf{\begin{table}[t]
    \begin{center}
    \caption{Data representation in the \textit{Certified Institution (CI)}\label{table:data_representation}}
    \begin{tabular}{ |c|c|c|c|c|c|c|c| }
        \hline
        \, & \multicolumn{6}{c|}{\textbf{Sequence}} & \textbf{Phenotype}\,\\
        \hline
         \textbf{Case} & \textbf{SNP\textsubscript{1}} & \textbf{SNP\textsubscript{2}} & \textbf{SNP\textsubscript{3}} & \textbf{SNP\textsubscript{4}} & \textbf{SNP\textsubscript{5}} & \, & \textbf{Cancer}\\
        \hline
        1 & AG & CC & TT & AG & CT & \ldots & Negative  \\
        2 & AA & CC & CT & AG & CT & \ldots & Negative  \\
        3 & AG & CT & CC & AA & TT & \ldots & Negative  \\
        4 & AG & CC & TT & AG & CT & \ldots & Negative  \\
        5 & GG & CT & TT & GG & CC & \ldots & Positive  \\
        6 & AA & CC & TT & GG & CC & \ldots & Positive  \\
        7 & AG & CT & CT & AG & CT & \ldots & Positive  \\
        8 & AA & CC & TT & GG & CC & \ldots & Positive  \\
        9 & GG & CT & CT & AG & CT & \ldots & Negative  \\
        10 & AG & CT & CT & AG & CT & \ldots & Positive  \\
        \hline
    \end{tabular}
    \end{center}
\end{table}}
\subsection*{Count Query}
Our objective is to securely execute count query operation, where \textit{researchers} want to know how many records in the database match a given query predicate. The number of SNPs in the query predicate is called the query size. We can formally define count query operation as follow:

\textit{\textbf{Definition 1}: Given a database \textit{D} and a query \textit{q}, count query can be defined as finding the number of tuples in $D$ which satisfies the predicate $\theta$ in $q$. If ${d_i}$ denotes one database tuple, the total count can be represented as: $\mid \{\forall i,\ d_i \in \textit{D}\ \mid d_i~satisfies~\theta\} \mid $}.
For example, let's consider the following query submitted by a \textit{researcher}:
\begin{Verbatim}[commandchars=\\\{\}]                                  
                SELECT COUNT(*) FROM Sequences
                    WHERE SNP\textsubscript{2} = `CC' AND SNP\textsubscript{3} = `TT' AND
                          SNP\textsubscript{5} = `CC' AND Cancer = 'Positive'
\end{Verbatim}
If we execute the above query on Table \ref{table:data_representation}, \textit{researchers} will receive $2$ as the answer of the query because only Case \# 6 and 8 match the query predicate. Determining this count value is an important step in genetic association studies. This value helps the \textit{researchers} to determine the association between genotype and phenotype. Computing the value of several statistical algorithms actually depends on this count value. These algorithm includes \textit{Pearson Goodness-of-Fit} or \textit{Chi-Squared Test}, \textit{Linkage Disequilibrium}, \textit{Estimation Maximization (EM)} and \textit{Cochran-Armitage Test for Trend (CATT)} \cite{Lauter2015:Springer}. 

\subsection*{Threat Model}
Our goal is that the \textit{CS} does not learn anything about the shared genomic data and both the \textit{CI} and the \textit{CS} learn nothing about the query performed by the \textit{researchers}. Furthermore, we want to ensure that the \textit{researchers} do not infer any information from the data. We assume the \textit{CI} to be a trusted entity as it is responsible for the generation and encryption of the \textit{index tree}. The \textit{CI} can verify the identity of the individuals or organizations who apply for the access of the data before giving them the keys. This role of verification performed by the \textit{CI} can be considered as the same as the \textit{Data Access Committee (DAC)} of NIH \cite{Paltoo:2014:Nature}. 

In our setting, we assume that the \textit{CS} is \textit{semi-honest}, also known as \textit{honest but curious} \cite{HazayLindell2010}. Such an adversary follows the protocol as specified and does not try to misrepresent any information about the contents. However, it may gather any statistical or other information regarding input, output or the computation during or after the protocol execution. Therefore, we require the view of each entity during protocol execution not to disclose any information. Thus, we assume that neither the \textit{data owner}, the \textit{CI}, nor the \textit{CS} has any motivation to behave maliciously in the desire to generate incorrect output.

Our method is also designed based on the the following assumptions: 
\begin{itemize}
\item We assume that the \textit{CI} does not collude with the \textit{CS} and \textit{CS} also does not collude with the \textit{researchers}. This is an essential requirement for guaranteeing query privacy.
\item We assume that the system works correctly, that is \textit{researchers} receive the correct keys from the \textit{CI}.
\end{itemize}

\section{Cryptographic Background}
In this section, we provide some relevant background information related to the cryptographic schemes we opt for our proposed framework. 

\subsection*{Yao's Garbled Circuit}
Yao's garbled circuit (Yao's protocol \cite{yao1986generate}) supports secure two-party computation against semi-honest adversaries. Let, two parties Alice ($A$) and Bob ($B$) wish to compute a function, $f(x, y)$ where $x$ and $y$ denote their respective inputs. The protocol evaluates the function $f$ through a Boolean circuit. The total number and kind of Boolean circuits necessary to calculate $f(x, y)$ depends on the function $f$. After running the protocol, both $A$ and $B$ learn the output of $f(x, y)$ but neither of them learn about the input of the other.

The working procedure of garbled circuit goes as follows. $A$ (known as the garbler) constructs a garbled version of the function $f(x,y)$ and sends it to $B$ (known as the evaluator) along with the garbled input of $A$. We denote this input as $I_g$. After receiving the circuit, the evaluator ($B$) runs a 1-out-of-2 oblivious transfer (OT) protocol \cite{rabin2005exchange}, to obliviously get the garbled circuit input values corresponding to its private input $I_e$. Through OT protocol $B$ learns only what he needs, and $A$ have no idea what $B$ learned. Using $I_g$ and $I_e$, the evaluator can calculate $f(x,y)$. 

\subsection*{Pailler Cryptosystem}
We use a semantically secure additive homomorphic encryption scheme called \textit{Paillier Cryptosystem} \cite{Paillier:1999:PCB} to encrypt the data and utilized its homomorphic properties to execute count query. A key generation algorithm produces a pair of keys: a secret key, ${sk}$ and a public key, ${pk}$. The public key and the secret key are used for encryption and decryption purposes respectively. Our encryption scheme is probabilistic; which means that if the same message is encrypted twice, we get two different ciphertexts. If we encrypt a messages $m$ twice and get two ciphertexts $c_1 = \xi_{pk}(m)$ and $c_2 = \xi_{pk}(m)$, then $c_1 \neq c_2$ and $\xi_{sk}(c_1) = \xi_{sk}(c_2) = m$.

\textit{\textbf{Homomorphic Properties:}} Assume that two messages $m_1$ and $m_2$ are encrypted using the same public key $pk$ and $k$ is a constant number. Then \textit{Paillier Cryptosystem} \cite{Paillier:1999:PCB} supports following homomorphic properties:
\begin{itemize}
\item If we multiply two ciphertexts, after decryption we will get the sum of their corresponding plaintexts.
\begin{align}
    \xi_{sk}(\xi_{pk}(m_1) \cdot \xi_{pk}(m_1)\ mod\ n^2)\ =\ m_1 + m_2 \ mod\ n
\end{align}
\item If we raise the power of a ciphertext to constant $k$, after decryption we will get the product of the corresponding plaintext and the constant.
\begin{align}
    \xi_{sk}(\xi_{pk}(m_1)^k\ mod\ n^2)\ =\ km_1\ mod\ n
\end{align}

\end{itemize}

\section{Methodology}
In this section, we present our proposed model. At first, \textit{CI} creates an \textit{index tree} from the datasets it receives from the \textit{data owners}, encrypts it and then sends it to \textit{CS}. The \textit{CS} uses this encrypted \textit{index tree} to execute queries. 

\subsection*{Preprocessing}

\begin{figure}[t]
\begin{subfigure}{0.3\textwidth}
  \centering
  \includegraphics[width=.30\linewidth]{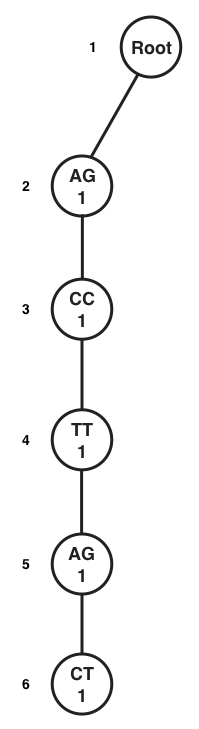}
  \caption{After the insertion of the first record}
  \label{fig:seq1}
\end{subfigure}%
\begin{subfigure}{.35\textwidth}
  \centering
  \includegraphics[width=.48\linewidth]{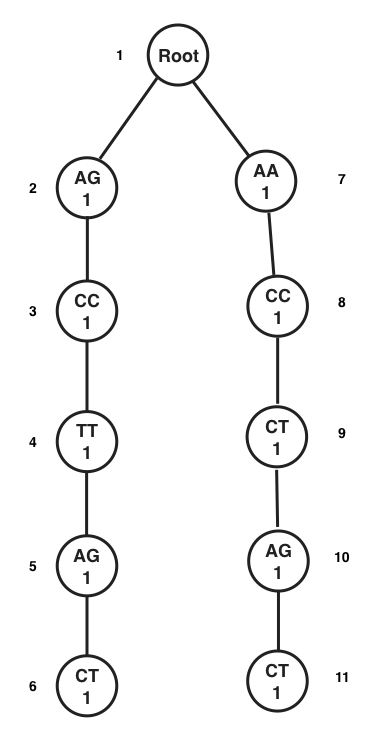}
  \caption{After the insertion of the second record}
  \label{fig:seq2}
\end{subfigure}
\begin{subfigure}{.35\textwidth}
  \centering
  \includegraphics[width=.67\linewidth]{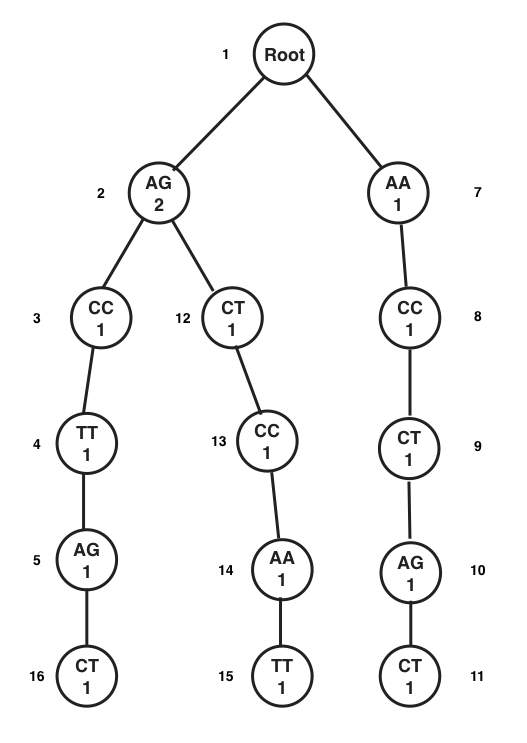}
  \caption{After the insertion of the third record}
  \label{fig:seq3}
\end{subfigure}
\caption{Different states during the generation of \textit{index tree}}
\label{fig:building_index_tree}
\end{figure}

\begin{figure}
\centering
    \includegraphics[width=3.6in]{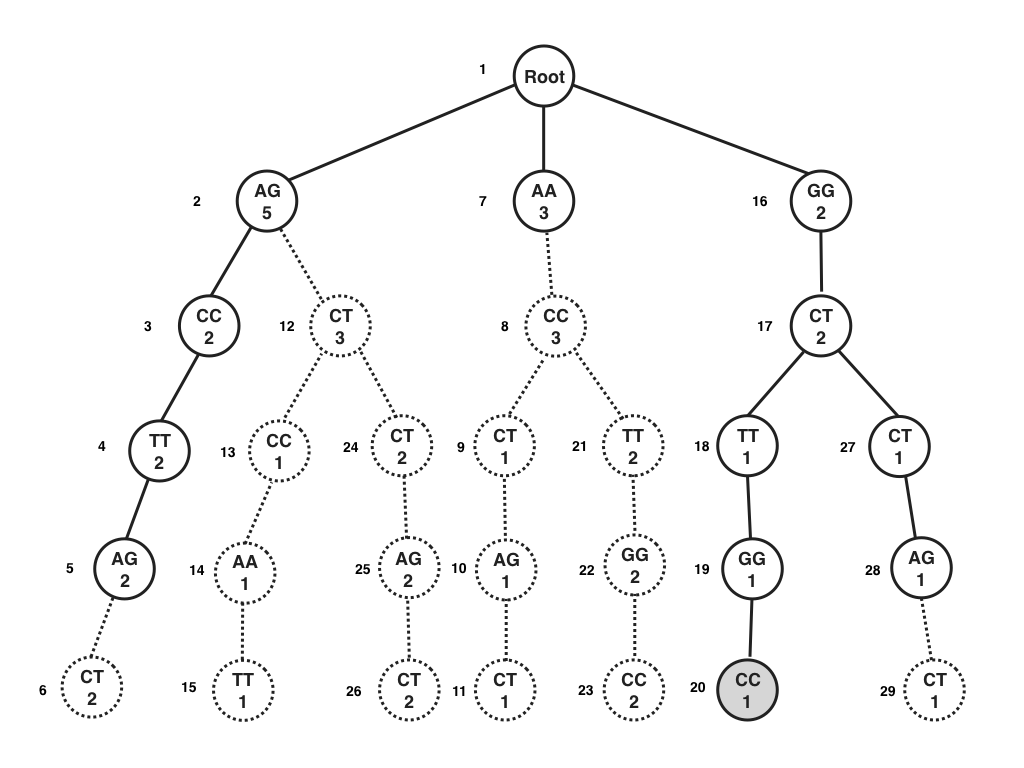}
\caption{\textit{Index tree} for Table \ref{table:data_representation}.}
\label{fig:prefix_tree}
\end{figure}
\textit{\textbf{A. Building Index Tree:}}
When the \textit{CI} receives the data from the \textit{data owners}, it first creates a search tree, $T$ which we call \textit{index tree}, using the SNPs from the database $D$. There is only one such tree in our system. After the creation of the tree, for the records from additional \textit{data owners}, \textit{CI} only needs to create or update the nodes in the $T$. For each record $d_i$ in the database, \textit{CI} encodes each SNP as:\\\\
\centerline{$d_i^j = k:1\leq$$i\leq$$\mid$$D$$\mid$; $1\leq$$j\leq$$\mid$$d_i\mid$; $1\leq$$k\leq$16}\\\\
Here, $\mid$D$\mid$ = number of records in the database and $\mid$$d_i$$\mid$ = number of columns in each record. Then, \textit{CI} checks if a node containing that SNP and SNP identifier is already in the tree or not. If not, then the \textit{CI} creates a new node for that SNP. Otherwise \textit{CI} just updates the corresponding existing node. Each node of $T$ contains:

\begin{enumerate}[label=\alph*)]
\item $sid$: the unique identifier for a SNP which occurs at a particular position in the genome.
\item $val$: the actual SNP , which is encoded as $\{1, 2, ..., 16\}$ for each of the 16 possible sequences. In Figure \ref{fig:building_index_tree} and \ref{fig:prefix_tree}, we have shown the actual SNP only for understanding purpose.
\item $count$: the number of times a SNP occur in that position.
\item $list$: the list of children each node contains (not shown in Figures \ref{fig:building_index_tree} and \ref{fig:prefix_tree}).
\end{enumerate}
We denote a node as $\sigma$ and represent as, $\sigma$($sid$, $val$, $count$, $list$). The tree $T$ is generated in the following way:

At first there is only one node in the tree which is the root node. Beginning from this root node, for each of the records in the database we start creating new nodes in $T$. We denote a record as $d_i^j\in D$ where $i$ indicates the record number and $j$ indicates the column number. For each $d_i^j$, the child of the root node is the corresponding first column $d_i^1$, the child of the node containing $d_i^1$ is the second column $d_i^2$ and we continue to create the tree in this way. So, in the \textit{index tree} the data from the first column is always on level $1$, data from the second column is on level $2$ and so on.

\textit{\textbf{Example 1:} The generated tree, $T$ after the insertion of first record $d_1$ from Table \ref{table:data_representation} is shown in figure \ref{fig:seq1}. Here, the first column, $d_1^1=AG$ is inserted as the child of the root node. The second column, $d_1^2=CC$ is inserted as the child of the node containing $d_1^1$ and so on. Each SNP occurs only once in the first record. So, each node contains the count value $1$. We can represent node \# 2 as $\sigma_2$(\textit{SNP}\textsubscript{1}, $AG,\ 1,\ \langle CC \rangle$).}

Now while inserting the second record, $d_2$ for each of the columns we first check whether the current column has already been inserted into a node in the corresponding level of $T$. If it has been inserted, we just increment the count value. Otherwise, we create a new node in that level to store $d_2^j$.

\textit{\textbf{Example 2:} Continuing from Example 1. The tree, $T$ after the insertion of second record $d_2$ is shown in Figure \ref{fig:seq2}. The first column for the second node $d_2^1$ is $AA$. We check if any existing node in $T$ already contains this SNP at level 1. Here, the root node has only one child $AG$. So, we create a new node and insert $AA$ as the child of the root node at level 1 and the following columns are added in the above mentioned way. For the third record, the first column $d_3^1=AG$ has already been inserted at level 1. So, we increment the value of $count$ at node 2. Now for second column, $d_3^2$ there is no child node of node \# 2 which contains $CT$. So, we create a new child node of node \# 2 at level 2 and then add the remaining columns similarly.}

Figure \ref{fig:prefix_tree} represents the \textit{index tree} containing all the records from Table \ref{table:data_representation}. All the nodes belonging to the same level represent a SNP all of which occur at a particular position of a genome which are actually represented as columns in Table \ref{table:data_representation}. Each node in the tree $T$ except the root node contains a value from a column. If there are $\theta$\textsubscript{n} number of columns in the database $D$, then the height of the index tree $T$ will be $\theta$\textsubscript{n}.

\begin{algorithm}[t]
  \caption{Algorithm for building tree from SNP sequence}\label{build}
  \textbf{Intput:} Root node and the database\\
  \textbf{Output:} This algorithm will return a index tree, $T$
  \begin{algorithmic}[1]
    \Procedure{\textit{BUILD-TREE}}{$r, D_i^j$}\Comment{r is the root node and D is the database}
    \For{each $D_i$}
        \State $parent\ \gets r$
        \State $a\gets \pi$($D_i^j$)\Comment{create new node for each $D_{i,j}$}
        \For{each $n \in$ $T$}
            \If{$a.sid\ =\ n.sid$ \textbf{and} $a.val\ =\ n.val$}\Comment{For each node we only match the $sid$ and $val$} 
                \State $n.$$count$$++$
                \State $parent\ \gets n$
            \Else
                \State $a.count\ \gets 1$
                \State $a.sid\ \gets \ \rho[j]$\Comment{$\rho[j]$ is a function which returns corresponding $sid$}
                \State $parent.addChildren(a)$
                \State $parent\ \gets a$
            \EndIf
        \EndFor
    \EndFor
    \State \textbf{return} $T$\Comment{The generated Tree}
    \EndProcedure
  \end{algorithmic}
\end{algorithm}

Algorithm \ref{build} provides pseudocode for building the \textit{index tree}. The building cost of the tree is $\mathcal{O}(mn)$ where, $m$ = number of records in the database and $n$ = number of different SNPs in the sequence. The features of this \textit{index tree} can be listed as:

\begin{itemize}
\item If we traverse one node at each level starting from the root node to a leaf node, we get different SNP sequences belonging to the same record in the database. For example, if we consider first record of Table \ref{table:data_representation}, the SNPs of this record are represented in the nodes 1, 2, 3, 4, 5 and 6. At each level, along with the SNP sequence, we also store the number of times that SNP sequence appears in that particular position of a genome. For example, in Figure \ref{fig:prefix_tree}, for SNP\textsubscript{1}, $AG$ occurs 5 times, $AA$ occurs 3 times and $GG$ occurs 2 times. Considering $AG$ as parent node for level 2, $CC$ occurs 2 times --- in this way all the nodes are created in $T$ with each SNP and the number of their occurrence.
\item We can again reconstruct the original database record by traversing the corresponding nodes of $T$.
\item For the addition or removal of records, we do not need to regenerate the tree, we can simply update or delete the data stored at each node. 
\item Unique SNP values at a particular level create new nodes and the following SNPs are added as the children of that node.
\item One noticeable characteristic of this tree is that if multiple predicates are involved, i.e. more than one SNP sequences are present in the query, then the resulting count value is equal to the value of the count stored at the node which matches the predicate located at the deepest level of the tree. So, if the \textit{researcher} is interested in SNP positions $x$, $y$ and $z$, and the position of $x$, $y$ and $z$ are such that $x < y < z$, then the count value is the value stored at node that represents the SNP sequence at position $z$. For example, consider the following query:
\begin{Verbatim}[commandchars=\\\{\}]                                  
                SELECT COUNT(*) FROM Sequences
                    WHERE SNP\textsubscript{1} = `GG' AND SNP\textsubscript{1} = `TT' AND
                          SNP\textsubscript{5} = `CC' AND Cancer = 'Positive'
\end{Verbatim}
Here, the value of count is 1 and it is the value that is stored at node that represents SNP\textsubscript{5} as this node is actually located at the deepest level of the tree among the nodes that matches the query.

\end{itemize}

\textit{\textbf{B. Encrypting the Index Tree:}}
After building the \textit{index tree} \textit{T} from the database, \textit{CI} encrypts the \textit{index tree} and then sends the encrypted version of \textit{T} to the \textit{CS}. The detailed process can be elaborated as:
\begin{itemize}
    \item \textit{Generate Keys:} The \textit{CI} generates a key pair ($pk,sk$) for a semantically secure additively homomorphic encryption scheme (\textit{Paillier Cryptosystem} \cite{Paillier:1999:PCB}) which consists of the following algorithms:
    \begin{itemize}
        \item \textit{KeyGen}: a key generation algorithm which generates a key pair ($pk,sk$) where $pk$ is the public key and $sk$ is the secret key.
        \item \textit{Enc:} an encryption algorithm which takes input a message $m$  and encrypts it using the public key $pk$. This is denoted as $\xi\textsubscript{pk}(m)$.
        \item \textit{Dec:} a decryption algorithm which takes input a ciphertext, $c$ and decrypts it using the secret key, $sk$. This is denoted as $\xi\textsubscript{sk}(c)$. Note that these encrypted records are not used in the search operation.
    \end{itemize}

    \item \textit{Encrypt the Index Tree:} \textit{CI} uses the public key, $pk$ to encrypt all the nodes in \textit{T}. To make the overall search process fast enough while maintaining the security of the system, it only encrypts the sensitive attributes in each node. For each node $\sigma$ in \textit{T}, it does $\xi\textsubscript{pk}(\sigma)$.
    After the encryption, each node is like $\sigma$($sid$, $\xi\textsubscript{pk}$\textit{(val)}, $\xi\textsubscript{pk}$($count$), $list$). We represent the encrypted database as $\widetilde{T}$.
    \item \textit{Share:} Finally, \textit{CI} sends ($pk$, $\widetilde{T}$) to the \textit{CS}. \textit{CI} also shares the key pair ($pk, sk$) with the \textit{researchers}.
\end{itemize}

\subsection*{Encryption of Query} 
The \textit{researchers} know about the format of the query he is allowed to perform. Once \textit{CI} sends ($pk, \widetilde{T}$) to the \textit{CS}, the \textit{researchers} can execute his query on $\widetilde{T}$ stored in \textit{CS}. He encrypts his query $q$ as $\xi_{pk}(q)$. Here, for the computation purpose, only $val$ is encrypted and $sid$ is kept in plaintext. So, we can represent the encrypted query as $\phi(sid, \xi(val))$. For example, the encrypted query is actually like:
\begin{Verbatim}[commandchars=\\\{\}]                                  
                SELECT COUNT(*) FROM Sequences
                    WHERE SNP\textsubscript{2} = `+a=#?h' AND SNP\textsubscript{4} = `z@0ux&*' AND
                        Cancer = '#?h\$ir*q!\%'
\end{Verbatim}

\subsection*{Searching on Index Tree}

\begin{algorithm}[t]
  \caption{Algorithm for searching SNP sequence in the tree}\label{search}
  \textbf{Input:} Root of encrypted \textit{index tree} and list of SNP identifiers in query\\
  \textbf{Output:} Resulting count value of the query
  \begin{algorithmic}[1]
    \Procedure{\textit{SEARCH-TREE}}{$r, s$}\Comment{$r$ is the root node and $s$ is the $sid$}
    \State $a\gets$ {$r.getChildren()$}
    \State $count\gets \xi$\textsubscript{pk}($0$)
    \While{$a\not=\{\phi\}$}
        \State $b\gets$ $a$.\textit{pop()}
        \If{$b.sid = s$}
            \State $c\ \gets\ Rand()$
            \State $d\gets\xi$\textsubscript{pk}($c$)
            \If{$\xi$\textsubscript{sk}($d$) - $i$ = $c$}\Comment{$i$ is the mapped SNP sequence value}
                \State $count\gets count + b.getCount()$\Comment{The equality checking is done using garbled circuit}
            \EndIf
        \EndIf
    \EndWhile\label{buildendwhile}
    \State \textbf{return} $count$\Comment{The count value}
    \EndProcedure
  \end{algorithmic}
\end{algorithm}

Our system supports the count operation. The search process starts with the \textit{researcher} sending the encrypted query $\phi$ to the \textit{CS}. The \textit{CS} needs to execute $\phi$ on $\widetilde{T}$ and find the number of records which matche the SNPs in the query predicate. For this, it requires to perform search operation on $\widetilde{T}$ and find the intended nodes which contain the $count$ values for the corresponding $sid$s. 

The main idea is to match the value of $val$ stored in the intended nodes (the $sid$ of these nodes matches with the $sid$ of $\phi$) which we denote as $val_n$ with the corresponding value of $val$ in the \textit{researcher's} query which we denote as $val_q$. If they match, \textit{CS} traverses the children of that node. This process continues until \textit{CS} finds all the nodes for the corresponding query or \textit{CS} finishes searching all the nodes of $\widetilde{T}$. As both the $val_q$ and \textit{val\textsubscript{n}} are encrypted and the encryption scheme we use is probabilistic, \textit{CS} cannot determine whether those values matche or not. The \textit{CS} can send the encrypted value of $val_n$ to the \textit{researcher} and as they have the secret key, they can decrypt $val_n$ and check the equality. But the problem of this approach is that the \textit{researchers} would be able to determine the structure of the tree using multiple query operations.

To enable search in this scenario while ensuring less information leakage to the \textit{researcher} and the \textit{CS}, we execute an interactive protocol between them to check this equality. This equality checking is basically done using garbled circuits. The \textit{CS} and \textit{researchers} compute this circuit via secure computation for each of the nodes which matches the value of $sid$ in the $\phi$. Here the \textit{researcher} is the garbler and \textit{CS} is the evaluator. Only the evaluator will know the output of the computation. As the $val_n$ is encrypted, this value can be decrypted into the circuit before checking the equality, but this process is computationally expensive \cite{nikolaenko2013privacy}.

We choose to use random mask to avoid this decryption inside the garbled circuit. The idea is to use the additive mask to obscure the input of \textit{CS} as the homomorphic property allows addition over encrypted data. We refer the additive mask we use as $noise$ and denote it as $\mu$. After the addition of the $noise$, the encrypted masked value of $val_n$ is:
\begin{align}
    \widetilde{\delta}\ =\ \xi_{pk}(val_n)\ +\ \xi\textsubscript{pk}(\mu)
\end{align}
Here, $\mu \in \mathcal{M}$ where $\mathcal{M}$ is the message space and $\mu$ is random. \textit{CS} then sends the resulting obscure value $\widetilde{\delta}$ to the \textit{researcher}. \textit{Researcher} get the masked value after the decryption as:
\begin{align}\label{masked_eqtn}
    \delta\ =\ \xi_{sk}(\widetilde{\delta})\ =\ val_n +\ \mu
\end{align}
\textit{Researchers} then subtract the corresponding value of $val_q$ from $\delta$ and get the noise as:
\begin{align} \label{noise_eqtn}
    \mu'\ =\ \delta\ -\ val_q
\end{align}
As $\mu$ is random, the \textit{researchers} will get random values for $\delta$ after decryption from Equation \ref{masked_eqtn}. As a result, though $\mu'$ is revealed to the \textit{researchers} from Equation \ref{noise_eqtn}, they will not be able to infer useful information from it.

The \textit{researcher} is the garbler of the circuit through which we check the equality. The input of the \textit{researcher} is $\mu'$. The input from \textit{CS} (evaluator) to this circuit is the actual noise it added, $\mu$. If the output of the circuit is true, then $\mu'\ =\ \mu$ which actually implies $val_n\ =\ val_q$. That means the SNP sequence in the \textit{researcher's} query matches with the SNP sequence in the database. Only \textit{CS} knows this output and \textit{CS} then continues traversing the children of that node. This process continues until \textit{CS} finds all the matched nodes for the corresponding query or \textit{CS} searched all the nodes of $\widetilde{T}$. 

\begin{figure}[t]
\centering
    \includegraphics[width=3.5in,height=4in]{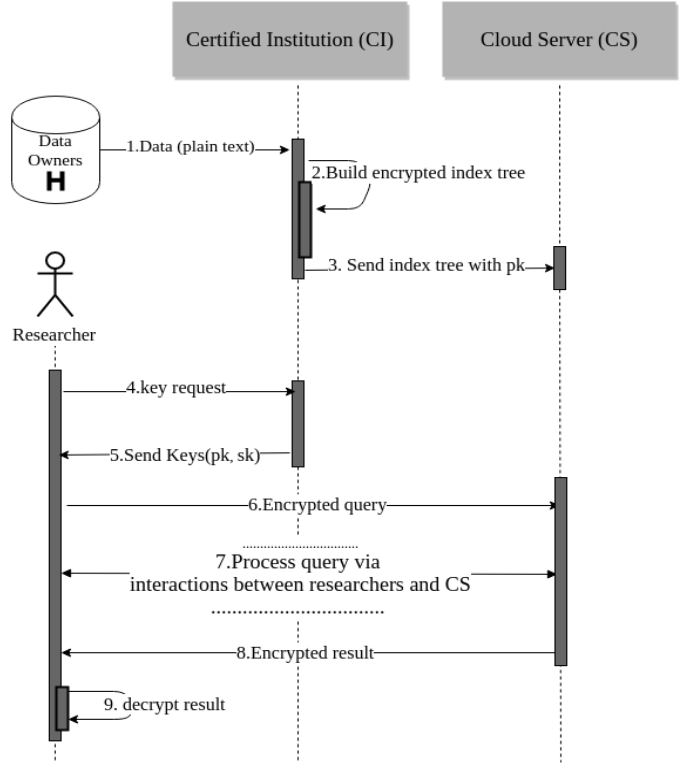}
\caption{Sequence diagram of our proposed model.}
\label{fig:seq_diagram}
\end{figure}

Algorithm \ref{search} provides the pseudo code for the search operation on $T$. Let $q$ be the query consisting of the SNPs the \textit{researchers} are interested in and $r$ be the root node of $T$. Let $sid$ be the SNP identifiers in $q$. Our search algorithm takes $r$ and $sid$ as input and returns the number of SNP sequences ($count$) that match the records in the database. Figure \ref{fig:seq_diagram} summarises each of the steps of our proposed method as sequence diagram.

\section{Security Analysis}
To determine the security of our system, we assume that the security of the system is compromised if the SNP sequences are revealed to any of the participants except the \textit{CI} as it is the trusted entity. We also consider the participants abilities to infer information in different stages of the system. The leakage profiles of different participants in our proposed model are given below --

\textbf{Leakage during the tree building and tree encryption phase:} \textit{CI} is only responsible for the  generation and encryption of the \textit{index tree} and is considered as a trusted entity. So the leakage to the \textit{CI} is none in this phase. The \textit{CS} cannot infer any information during this phase as it only gets the encrypted \textit{index tree}, $\widetilde{T}$.

\textbf{Leakage to \textit{CI} in each query:} The \textit{CI} is not involved at all during the query execution, it's only responsibility is to provide the key pair $(pk, sk)$ to the \textit{researchers}. The leakage to \textit{CI} during the query execution is none.

\textbf{Leakage to \textit{researchers} in each query:} The leakage to the \textit{researchers} is the final output which is the result of the query. \textit{Researchers} also know the noise value, $\mu'$ from Equation \ref{noise_eqtn} but $\mu'$ is a random number and uniformly distributed. Hence, the \textit{researchers} cannot infer anything from the value of $\mu'$. 
Note that we do not consider here any privacy leakage through the output. Such inference attack can be avoided using \textit{differential privacy} and has been studied extensively in the literature~\cite{dwork2014algorithmic}. 

\textbf{Leakage to \textit{CS} in each query:} The \textit{CS} can know the tree traversal path, that is all the nodes in $\widetilde{T}$ which are accessed during the query execution. Depending on the result of the query, the tree traversal pattern includes either the paths reaching the leaves, or the paths stop at some internal nodes. \textit{CS} can learn about the the SNP identifiers from a query but not the SNP sequences, because the SNP sequences are encrypted but the SNP identifiers are not. As the output of the circuit computation is only known to the \textit{CS}, it can know which node actually contains which SNP identifier. But as the SNP sequences and all other information stored in that node are encrypted, \textit{CS} cannot learn about any other values from that node.

\section{Experimental Results}
We have built a prototype of our privacy preserving system to evaluate its practicality and tested its performance on both real and synthetic datasets. The \textit{CS} and the \textit{CI} run on two different machines. Both of them were Intel Core i5 3.3 GHz processors with 8 GB RAM, running Ubuntu Linux 16.04. The source code is written in JAVA programming language. For the simulation purpose, we considered the users separately. 

We consider the following aspects in order to estimate the efficiency of our proposed method:
\begin{enumerate}
\item \textit{Data read and tree building time:} Time needed to process genomic database and build \textit{index tree}.
\item \textit{Tree encryption time:} Time needed to encrypt the \textit{index tree}.
\item \textit{Query execution time:} Time needed to execute a query submitted by the \textit{researchers}.
\item \textit{Communication overhead:} Bandwidth requirement between the evaluator (\textit{CS}) and garbler (\textit{researchers}) in order to process a count query.
\end{enumerate}

We also compare our proposed method with the methods proposed by Kantarcioglu \textit{et al.} \cite{Kantarcioglu2008IEEE} and Canim \textit{et al.} \cite{Canim2011IEEE}. Note that we can only compare their method for query execution time as our proposed method is different from them.

We have implemented the cryptography building block that were described in Section 4. We investigated different garbled circuit libraries such as \textit{FastGC} \cite{huang2011faster}, \textit{ObliVM-GC} \cite{liu2015oblivm}, \textit{JustGarble} \cite{bellare2013efficient} and used the \textit{FlexSC} \cite{wang2015circuit} library to implement the garbled circuits. We also used the \textit{Paillier Cryptosystem} \cite{Paillier:1999:PCB} to implement the \textit{homomorphic encryption}. 

To evaluate our system on real life dataset, we used the dataset available from the iDash competition 2015 \cite{iDash2015} where there are 311 different SNP sequences from 400 different participants divided into case and control groups. As the real-life dataset was not large enough to evaluate the scalability of our proposed model, we generated different synthetic datasets varying the number of records (between 10K to 50K) and number of SNPs (between 60 to 300) by randomly adding records to the iDash competition 2015 \cite{iDash2015} dataset. We experimented with five different query size that involve 10, 20, 30, 40 and 50 randomly selected SNP sequences. For each experiment, we executed 10 runs and averaged the result over the runs.

We organize the experimental analysis into four following parts to evaluate various perspective of our proposed framework.

\begin{figure}[t!] 
  \begin{subfigure}[b]{0.5\linewidth}
    \centering
    \includegraphics[width=0.85\linewidth]{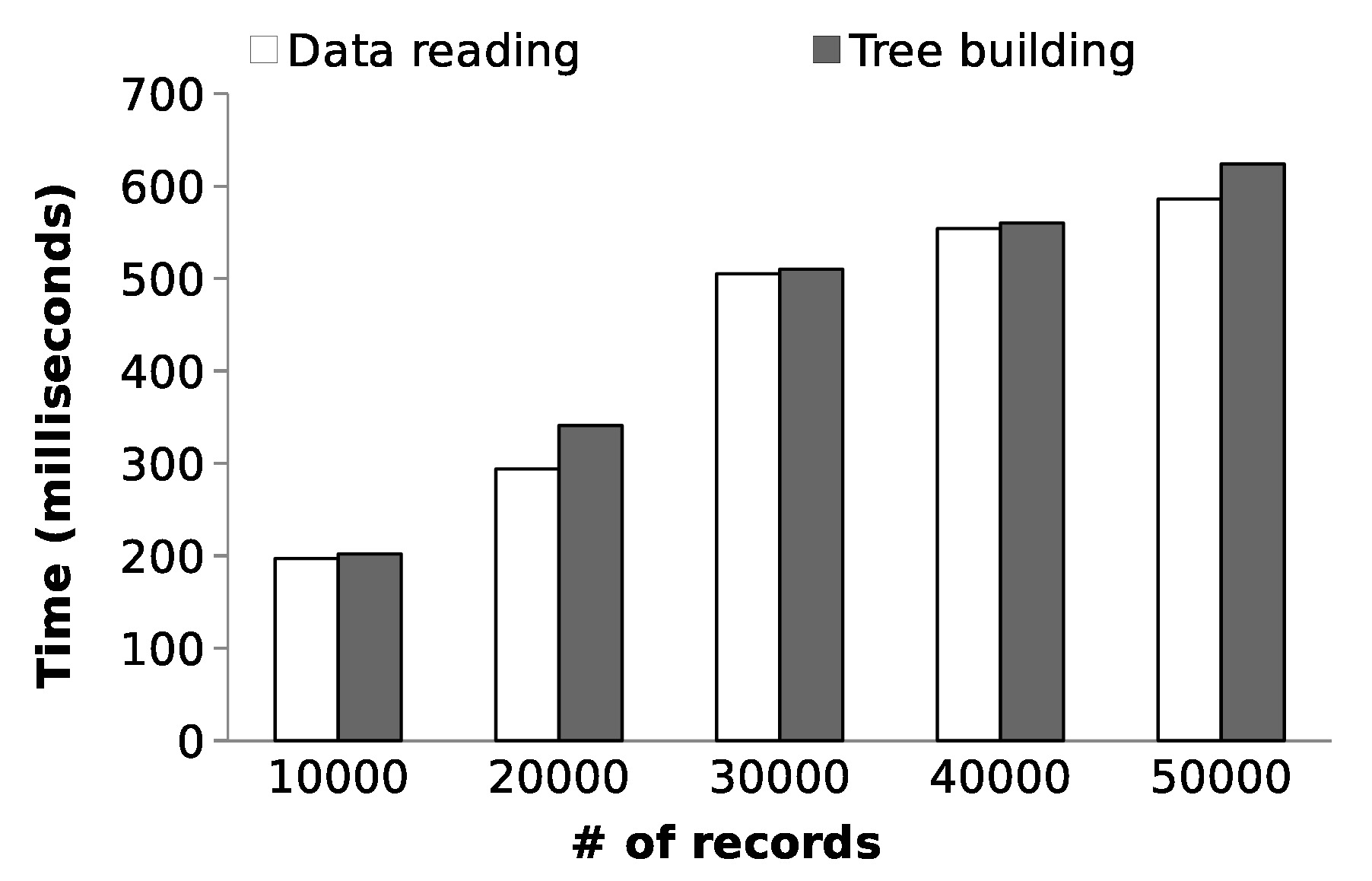} 
    \caption{Data read and \textit{index tree} building time.} 
    \label{fig:data_read_and_tree_build} 
  \end{subfigure}
  \begin{subfigure}[b]{0.5\linewidth}
    \centering
    \includegraphics[width=0.85\linewidth]{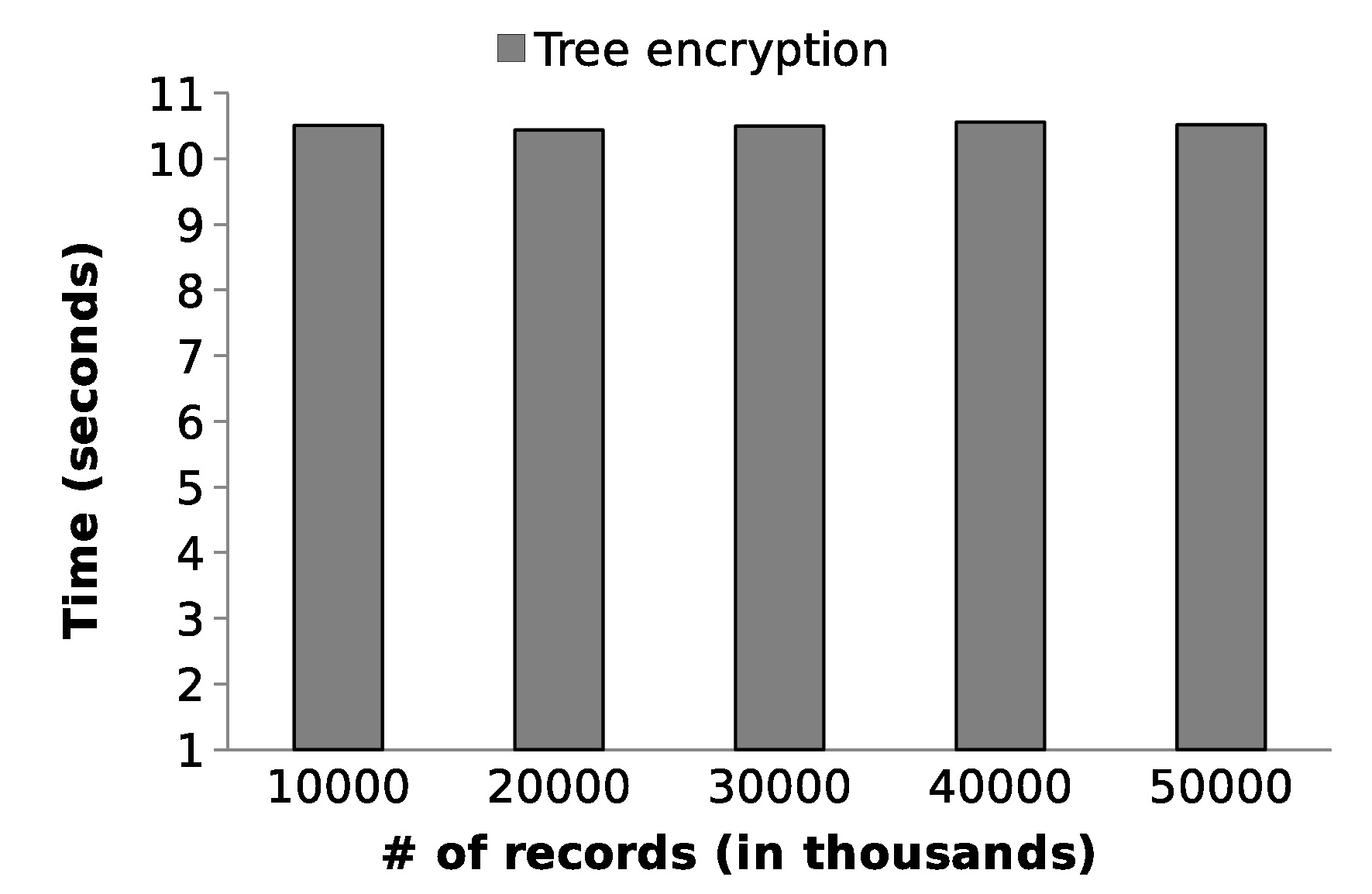} 
    \caption{Tree encryption time.} 
    \label{fig:tree_encrypt} 
  \end{subfigure} 
  \caption{Execution time of data read, \textit{index tree} building and tree encryption. Data read and \textit{index tree} building time linearly increases with the number of records. However, the tree encryption time is changes very little with the increase of number of records. This is because, only total number of SNPs in a record affects the depth of \textit{index tree}, more SNPs implies higher depth. Hence, the tree encryption time is similar regardless of the number of records in the dataset.}
  \label{fig:tree} 
\end{figure}

\begin{figure}[t!] 
  \begin{subfigure}[b]{0.5\linewidth}
    \centering
    \includegraphics[width=0.85\linewidth]{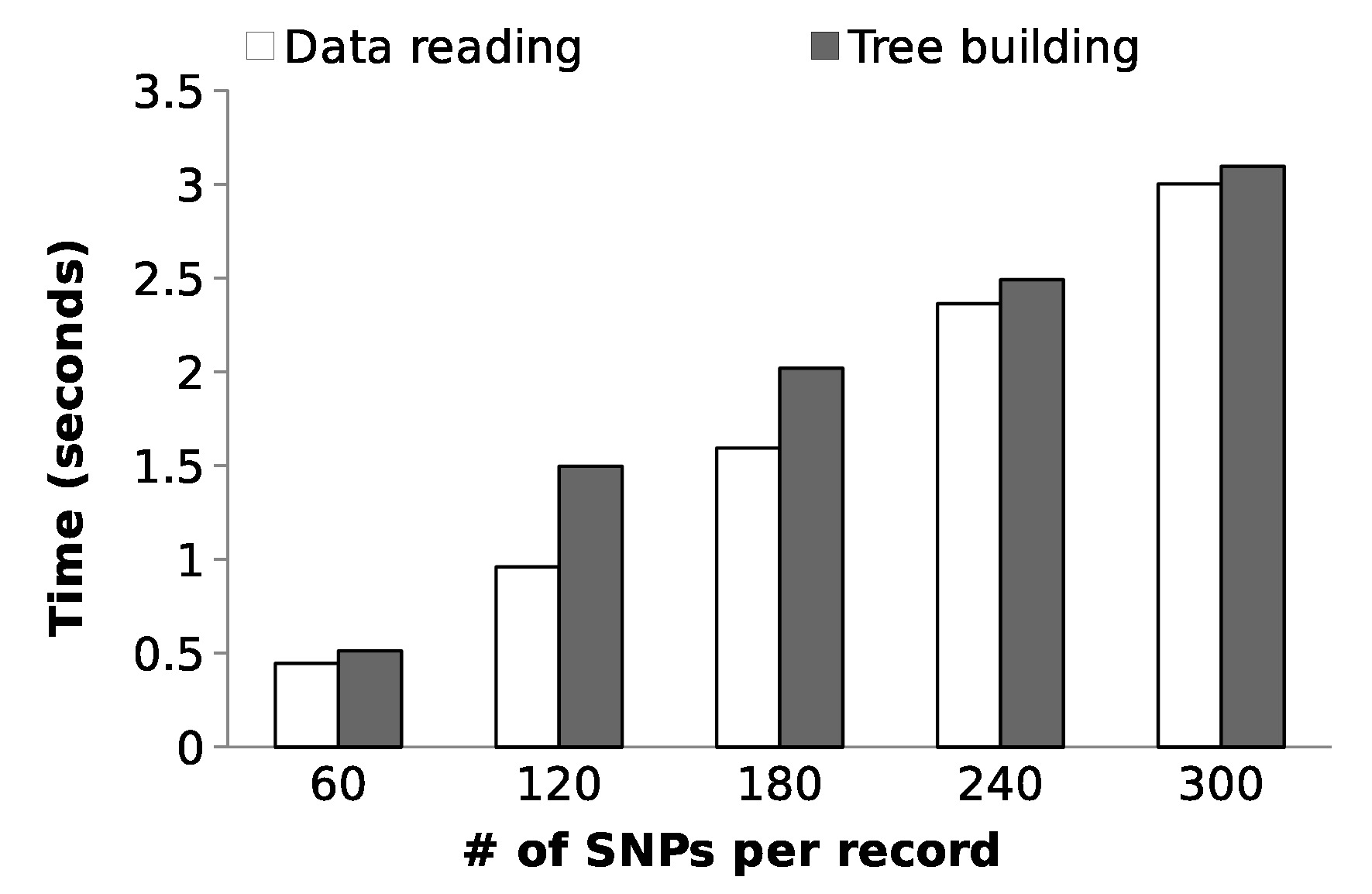} 
    \caption{Data read and \textit{index tree} building time.} 
    \label{fig:data_read_and_tree_build_col} 
  \end{subfigure}
  \begin{subfigure}[b]{0.5\linewidth}
    \centering
    \includegraphics[width=0.85\linewidth]{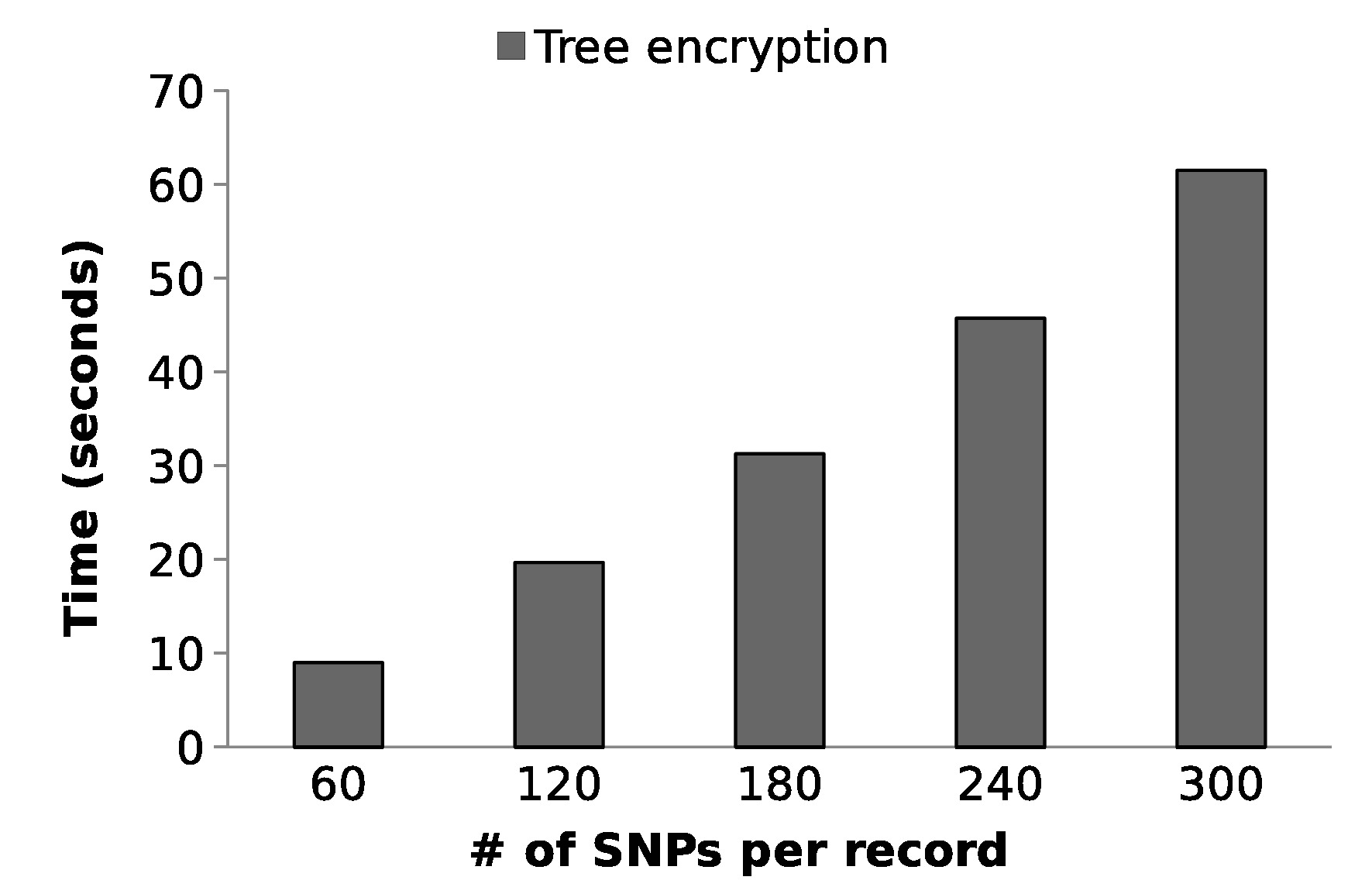} 
    \caption{Tree encryption time.} 
    \label{fig:tree_encrypt_col} 
  \end{subfigure} 
  \caption{Execution time of data read, \textit{index tree} building and tree encryption in increasing number of SNPs per records. Data read, \textit{index tree} building time, and tree encryption time linearly increases with the number of SNPS per records.}
  \label{fig:scalability_col} 
\end{figure}

\textbf{A. Data Read and Tree Building Time:} In order to determine the scalability of our system, we analyzed the time required for different datasets containing different number of records. Figure \ref{fig:data_read_and_tree_build} plots the time required for reading the data from the database and building \textit{index tree} using this data. Here we fixed the number of SNPs to 70 and increased the number of records from 10K to 50K. As expected, the time increases linearly with the increase of number of records. Thus, when we increase our number of records to 50000, the data read and \textit{index tree} building time increases to approximately 0.5 seconds and 0.6 seconds respectively.

\textbf{B. Tree Encryption Time:} Figure \ref{fig:tree_encrypt} plots the tree encryption time for datasets with different number of records. Each of the dataset containe 70 SNPs. Our experiments show that the increase in number of records do not significantly impact the encryption time. This is due to the fact that the encryption time depends on the depth of \textit{index tree} and depth of the tree in turn depends on the total number of SNP sequences in the dataset, not on the number of records. Thus, tree encryption time is almost similar regardless of the number of records in the dataset.
\begin{figure}[t!] 
  \begin{subfigure}[b]{0.5\linewidth}
    \centering
    \includegraphics[width=0.95\linewidth]{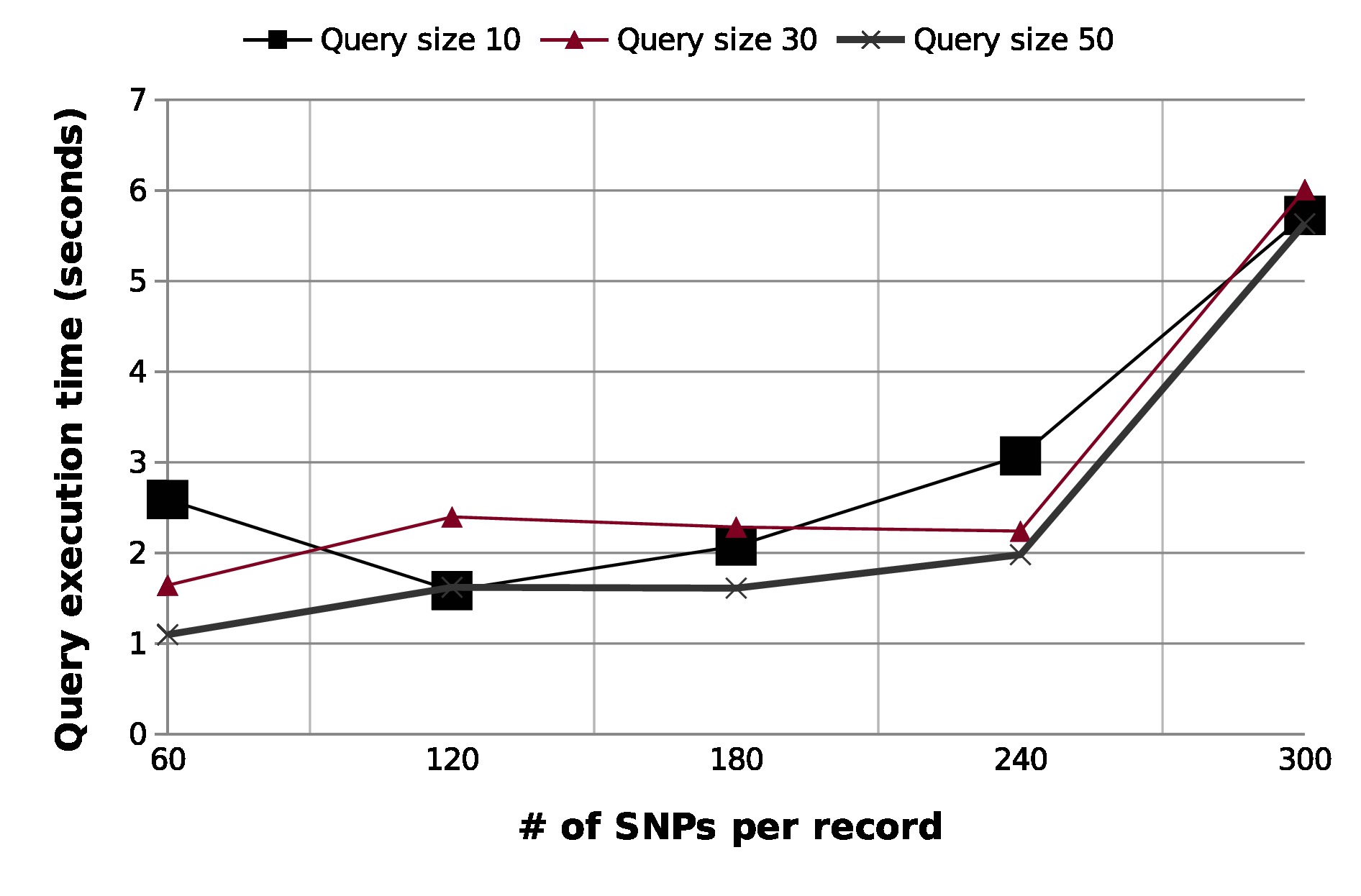} 
    \caption{Query execution time (number of records: 50K).} 
    \label{fig:query_eva} 
    \vspace{2ex}
  \end{subfigure}
  \begin{subfigure}[b]{0.5\linewidth}
    \centering
    \includegraphics[width=0.95\linewidth]{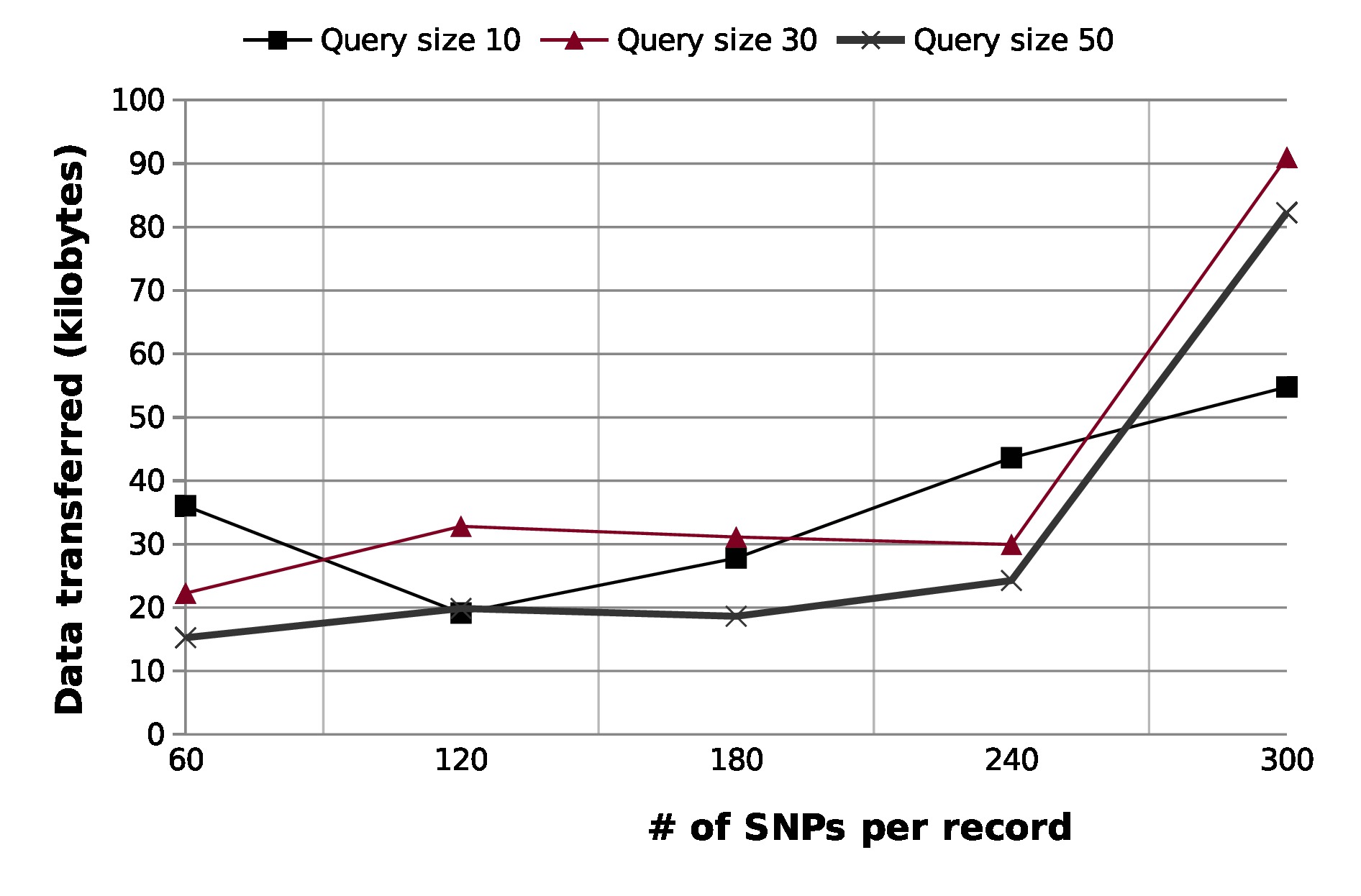} 
    \caption{Communication overhead (number of records: 50K).} 
    \label{fig:comm_overhead}
    \vspace{2ex}
  \end{subfigure}
  \begin{subfigure}[b]{0.5\linewidth}
    \centering
    \includegraphics[width=0.95\linewidth]{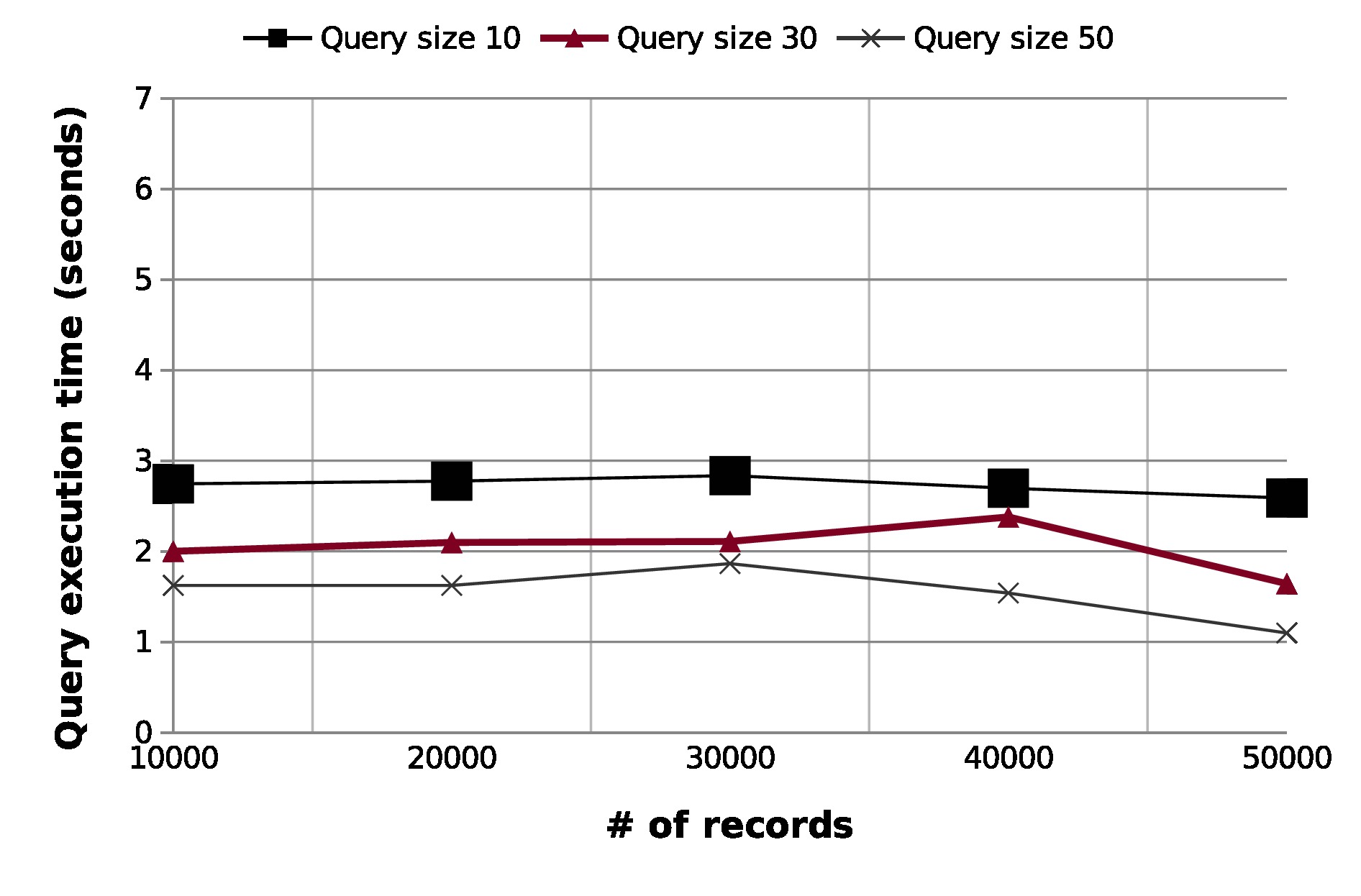} 
    \caption{Query execution time (number of SNPs per record: 60).} 
    \label{fig:query_eva1} 
    \vspace{2ex}
  \end{subfigure}
  \begin{subfigure}[b]{0.5\linewidth}
    \centering
    \includegraphics[width=0.95\linewidth]{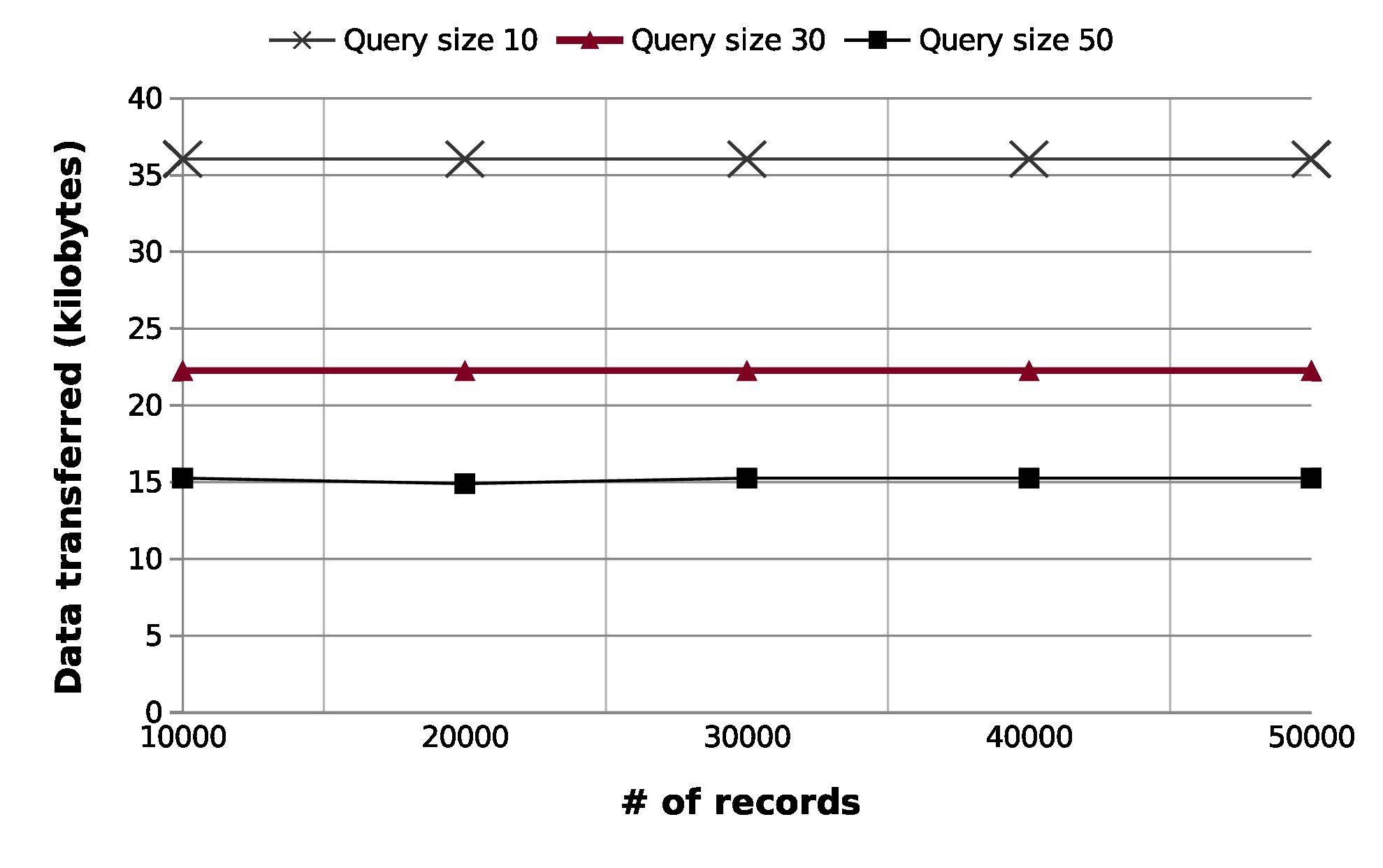} 
    \caption{Communication overhead (number of SNPs per record: 60).} 
    \label{fig:comm_overhead1}
    \vspace{2ex}
  \end{subfigure}
  \begin{subfigure}[b]{0.5\linewidth}
    \centering
    \includegraphics[width=0.95\linewidth]{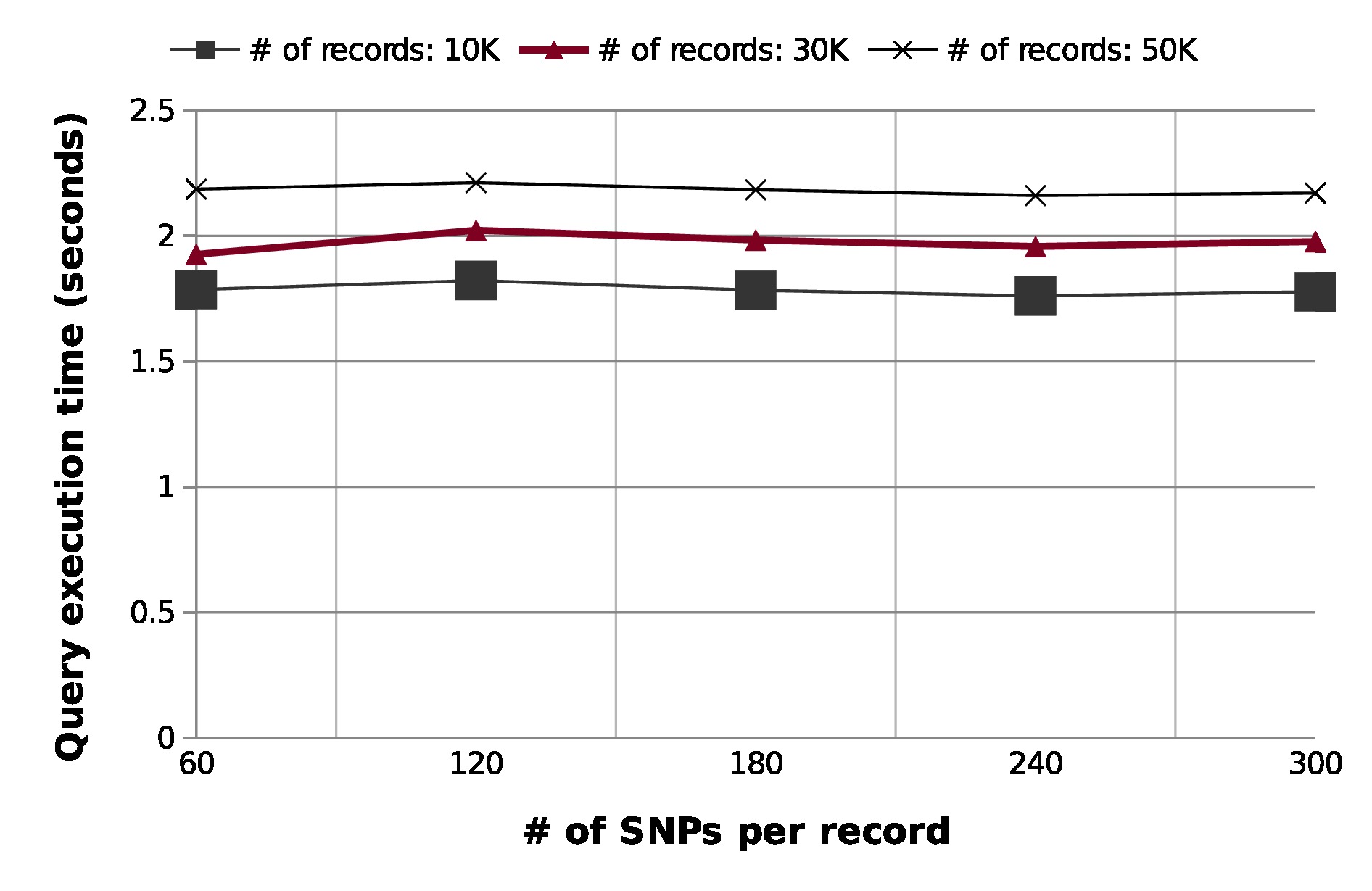} 
    \caption{Query execution time (query size: 30).} 
    \label{fig:query_eva2} 
    \vspace{0.5ex}
  \end{subfigure}
  \begin{subfigure}[b]{0.5\linewidth}
    \centering
    \includegraphics[width=0.95\linewidth]{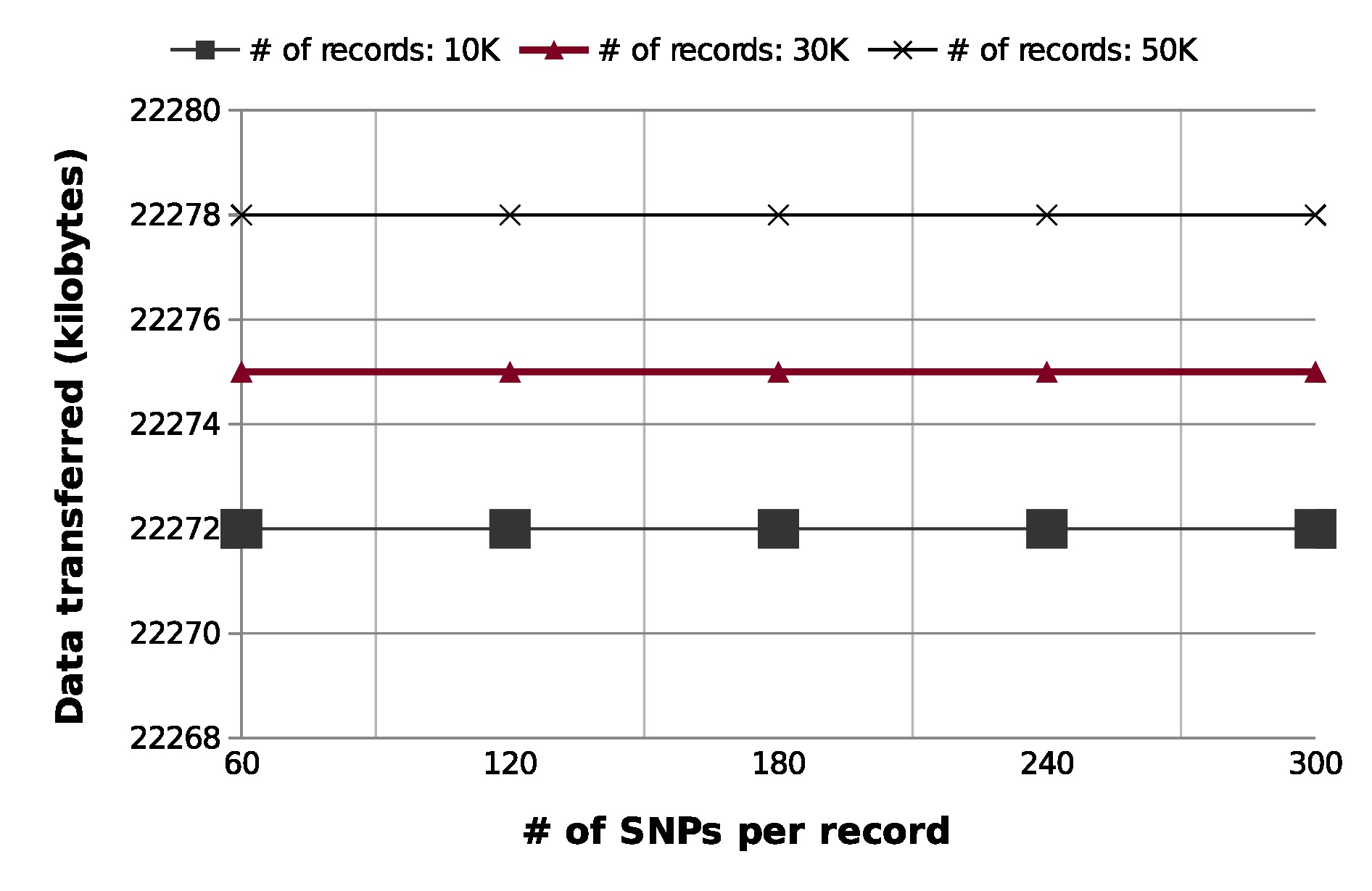} 
    \caption{Communication overhead (query size: 30).} 
    \label{fig:comm_overhead2}
    \vspace{0.5ex}
  \end{subfigure}
  \caption{Query execution time and communication overhead for count queries on datasets with different number of records, SNPs per record, and query sizes. In Figure \ref{fig:query_eva} and \ref{fig:comm_overhead}, we experiment with 50000 number of records while changing SNPs per record and query size. Note that increasing the number of query size or SNPs do not significantly affect on the query execution time or communication overhead. In Figure \ref{fig:query_eva1} and \ref{fig:comm_overhead1}, we change total number of records and query size while restricting 60 SNPs per record. We observe that query time and communication overhead slightly decreases as the query size increases. This is due to the fact that for a larger query size of an \textit{AND} query, traversing higher depth of the tree lessen the probability of getting many matches and vice versa. Figure \ref{fig:query_eva2} and \ref{fig:comm_overhead2} demonstrate query time and communication overhead for the fixed query size 30, while changing total number of records and SNPs per record. Note that query time and communication overhead do not significantly increases as the total number of records or SNPs per record increases.}
  \label{fig:query&comm} 
\end{figure}

In the next experiment we fixed the number of records to 50K and increased the number of SNPs from 60 to 300. As the number of SNPs increases, the total execution time also increases linearly. Figure \ref{fig:scalability_col} depicts the execution time of data read, \textit{index tree} building and tree encryption time for this experiment.

\textbf{C. Query Execution Time:} Figure \ref{fig:query_eva}, \ref{fig:query_eva1}, and \ref{fig:query_eva2} plots the query execution time based on three different parameters. These parameters are -- \textit{a}) number of records in the dataset, \textit{b}) number of SNPs in the dataset and \textit{c}) query size. In each of the experiment we retained one parameter fixed while changing the others. We note that the size of the query, size of the datasets or number of SNPs in the dataset do not significantly affect the query execution time. Rather, this time directly depends on how deep the levels of the \textit{index tree} need to be traversed and different traversing pattern of the \textit{index tree} depending on the query.

\textbf{D. Communication Overhead:} Figure \ref{fig:comm_overhead}, \ref{fig:comm_overhead1}, and \ref{fig:comm_overhead2} plots the amount of data transferred between the \textit{researchers} and the \textit{CS} during the evaluation of the garbled circuits. In these experiments again we considered three different parameters mentioned earlier and retained one parameter fixed while changing the others. Here we also observed that size of the datasets, number of SNPs in the dataset or the size of the query do not significantly affect the communication overhead. It actually depends on the tree traversal pattern based on the query.

\subsection*{Improvements Over Prior Approaches}
We compare the time required to execute a query by our proposed model with Kantarcioglu \textit{et al.}~\cite{Kantarcioglu2008IEEE} and Canim \textit{et al.}~\cite{Canim2011IEEE}. Though they did not build search index like us, we make this comparison to stress the efficiency of our proposed model in terms of query execution time. Also note that our calculated query execution time only includes the time required to encrypt and execute the query and then decrypt the final result. For the comparison purpose, we experimented on the datasets with similar number of records and columns used by Kantarcioglu \textit{et al.}~\cite{Kantarcioglu2008IEEE} and Canim \textit{et al.}~\cite{Canim2011IEEE}. The result of the comparison is listed in Table \ref{table:comparison}. A query consisting of 40 SNPs in a dataset of 5000 records, completed in approximately 30 minutes and 80 seconds by the models proposed by Kantarcioglu \textit{et al.}~\cite{Kantarcioglu2008IEEE} and Canim \textit{et al.}~\cite{Canim2011IEEE} respectively, whereas our proposed framework completes the similar query only in 1.7 seconds. Therefore, our protocol is about 900 times faster than Kantarcioglu \textit{et al.}~\cite{Kantarcioglu2008IEEE} and 40 times faster than Canim \textit{et al.}~\cite{Canim2011IEEE} in terms of query execution time. We also observe that, our improvement ratio is independent of the size of the dataset.

\begin{table}[t]
\vspace{.4cm}
    \begin{center}
    \begin{tabular}{ |c|c|c|c|c|c|c| }
        \hline
         \textbf{Method} & \textbf{Query size 10} & \textbf{Query size 20} & \textbf{Query size 30} & \textbf{Query size 40}\\
        \hline
        Kantarcioglu \textit{et al.}~\cite{Kantarcioglu2008IEEE} & 25 min & 27 min & 28 min & 30 min  \\
        Canim \textit{et al.}~\cite{Canim2011IEEE} & 20 sec & 40 sec & 60 sec & 80 sec  \\
        Our method & 2.4 sec & 2.7 sec & 1.5 sec & 1.7 sec  \\
        \hline
    \end{tabular}
    \caption{Comparison of count query execution time on 5000 records with query sizes between 10 to 40 SNPs.
    \label{table:comparison}}
    \end{center}
\end{table}
\subsection*{Summary}
Our experimental results on different datasets by varying the number of records, SNPs and query sizes can be summarized as:
\begin{itemize}
    \item Our method can effectively preserve both data privacy and data utility supporting large datasets by building an \textit{index tree}. We observe that the time required to read the data from the database and build \textit{index tree} using this data is linear (Figure \ref{fig:data_read_and_tree_build}). Also the tree encryption time does not have direct impact on the number of records (Figure \ref{fig:tree_encrypt}).
    \item By comparing with the existing solutions we have demonstrated that our proposed method is superior than the others in terms of query execution time.
    \item Our proposed framework is highly scalable for large datasets. 
\end{itemize}
These characteristics make our proposed method a promising framework for secure query search for biomedical data.

\section{Conclusion}
In this paper, we have presented a secure and efficient method for outsourcing genomic data. The proposed method constructs an \textit{index tree} from the aggregated genomic data and then outsources it to the third party cloud server. By employing a secure interactive protocol, the cloud server can traverse the nodes of the tree and execute count query operation. We have demonstrated that our model does not reveal any sensitive genomic data during the data processing as well as query execution phase. It is evident from our experiments on both real and synthetic datasets that our proposed model provides better performance than the existing solutions in terms of query execution time. However, in this research we have kept our work limited only to genotypes. We plan to extend our current method by incorporating phenotype information as well as supporting more complex operations along with the count query.

\section*{Acknowledgments}
We sincerely thank the reviewers for their insightful comments. The research is supported in part by the NSERC Discovery Grants (RGPIN-2015-04147) and University of Manitoba Startup Grant.

\makeatletter
\renewcommand{\@biblabel}[1]{\hfill #1.}
\makeatother

\bibliographystyle{unsrtnat_xq}
\bibliography{amia}

\end{document}